\documentclass[9pt,twocolumn,twoside]{opticajnl}
\journal{opticajournal} % use for journal or Optica Open submissions

\setboolean{shortarticle}{true}
\usepackage[utf8]{inputenc}
\title{Spatio-temporal coherent molding and retrieval of pulsed signals in  optical  waveguides}

\author[1]{Nika Durishvili}
\author[2,*]{Ramaz Khomeriki}
\author[3,4]{Vakhtang Jandieri}
\author[5]{Douglas H. Werner}
\author[4]{Daniel Erni}
\author[6]{Jamal Berakdar}

\affil[1]{Free University of Tbilisi, School of Mathematical and Computer Sciences, 0159 Tbilisi, Georgia}
\affil[2]{Physics Department, Javakhishvili Tbilisi State University, 3 Chavchavadze, 0128 Tbilisi, Georgia}
\affil[3]{Department of Electrical and Mechanical Engineering, Nagoya Institute of Technology, Gokiso-cho, Showa, Nagoya, Aichi 466-8555, Japan}
\affil[4]{General and Theoretical Electrical Engineering (ATE), Faculty of Engineering, University of Duisburg-Essen and Center for Nanointegration Duisburg-Essen (CENIDE), D-47048 Duisburg, Germany}
\affil[5]{Department of Electrical Engineering, The Pennsylvania State University, University Park, PA 16802, USA}
\affil[6]{Institut f\"ur Physik, Martin-Luther-Universit\"at, Halle-Wittenberg,  06099 Halle/Saale, Germany}

\affil[*]{khomeriki@hotmail.com}

\begin{abstract}
Optical waveguides are key elements for high fidelity, long distance optical communications. Coupled waveguide arrays allow for higher information density, steering the propagation direction, and for encoding  information. However, due to the mixing of relative phases for short pulses containing multiple waveguide-mode frequencies, a process for retrieving an encoded input state once these signals undergo coherent propagation remains elusive. A concept  is presented  to  extract with high fidelity the phase-encrypted input signal  from  spatio-temporal propagated states. As a realization, an array of coupled waveguides is suggested with the retrieval mechanism being realized by local phase shifts that  comply with the identified  retrieval concept. Three-dimensional full-wave electromagnetic simulations for broadband optical signals in coupled dielectric waveguides 
confirm the validity of the scheme and the high fidelity of information retrieval pointing  to
potential applications, for instance in ultrafast coherent coding and decoding of information imprinted on  pulse sequences.  
\end{abstract}

\setboolean{displaycopyright}{false} % Do not include copyright or licensing information in submission.

\begin{document}

\maketitle

\begin{figure}[h]
\includegraphics[width=\linewidth]{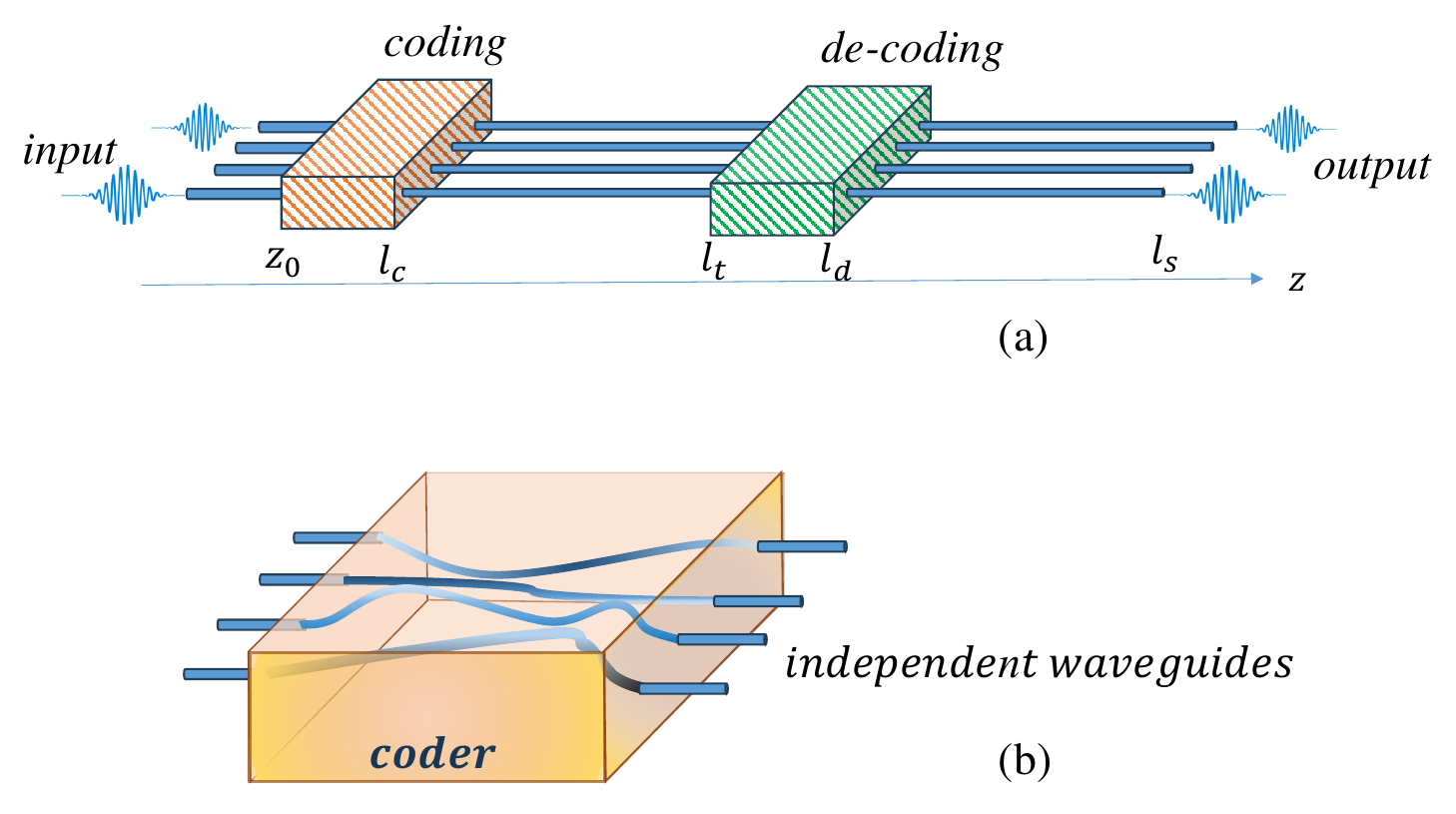}
\caption{{ (a) Schematics for coding (from $z = z_0$ to $z = l_c$) and decoding  (from $z = l_t$ to $z = l_d$) device. (b) The coder and decoder are composed of  the coupled waveguides. The pulses are injected in independent waveguides at $z < z_0$.  Coupling causes the power to randomly distribute at $z = l_c$, encrypting the input signal. The pulses are transferred  to the decoder at $z=l_t$ via independent  transmission lines. At $z = l_t$ the signal is decoded according to the scheme in the manuscript.} } \label{Pic}
\end{figure}

\noindent
{\it Introduction.}  Due to size quantization, high-quality  optical fibers are capable of hosting electromagnetic  long-lived eigenmodes at discrete frequencies and  particular  propagation (constant or)  wavevectors. This fact enables the use of optical fibers for low loss information processing and transfer \cite{Okamoto_2022,Kim_2024}. When the waveguides are located in close proximity such as in Fig. \ref{Pic}, coupling between the modes may occur and a locally injected signal tunnels coherently across the entire ensemble. After some propagation distance, it is therefore not possible to infer the characteristics of the injected signal from local measurements. 
Clearly, this may offer a way to code optical information but such an  encryption is not useful unless a scheme is developed to retrieve faithfully the encoded data. For swift communication and high level of encryption, broadband pulses are desirable but pose a challenge, as the recovery scheme must be capable of retracing in space and time coherent interacting waves  with multiple-frequencies. 
The issue of reverting a system evolution is also of general interest and is recurrently discussed  starting with the debate  between Loschmidt and Boltzmann \cite{LB1,LB2}. Since then, numerous  examples have been presented with the phenomenon of spin echo \cite{spin} being probably the most prominent one. Time reversal process has also been employed in the analysis of water surface waves \cite{water1,water2}, Dirac lattices \cite{dirac}, and cold atoms in optical lattices \cite{ch}.\\ 

In optics, input signal identification has been addressed experimentally  \cite{szameit1,szameit2} by using  fragmented waveguides with well defined lengths,  which induces  appropriate  relative phase shifts  of the wave in  neighboring waveguides. Spatial recovery of  stationary beams is achieved but the  procedure  fails for short   pulses because   the phase shifts acquired by various  harmonics in the broadband pulse   are  different while the fragmented waveguides are tuned to a certain frequency. Alternative discussed mechanisms rely on an exchange of sublattices in binary waveguides \cite{longhi},  switching of  eigenstates \cite{peschel},  use of  zero-gap periodic systems \cite{pulse}, and { other signal transforming approaches  \cite{NiuNiuHuHuDuYuChu+2023+3737+3745,Teimourpour:16,Heinrich:12,ref1,ref2,ref3,ref4}}.

\begin{figure}[t]
\includegraphics[width=0.9\linewidth]{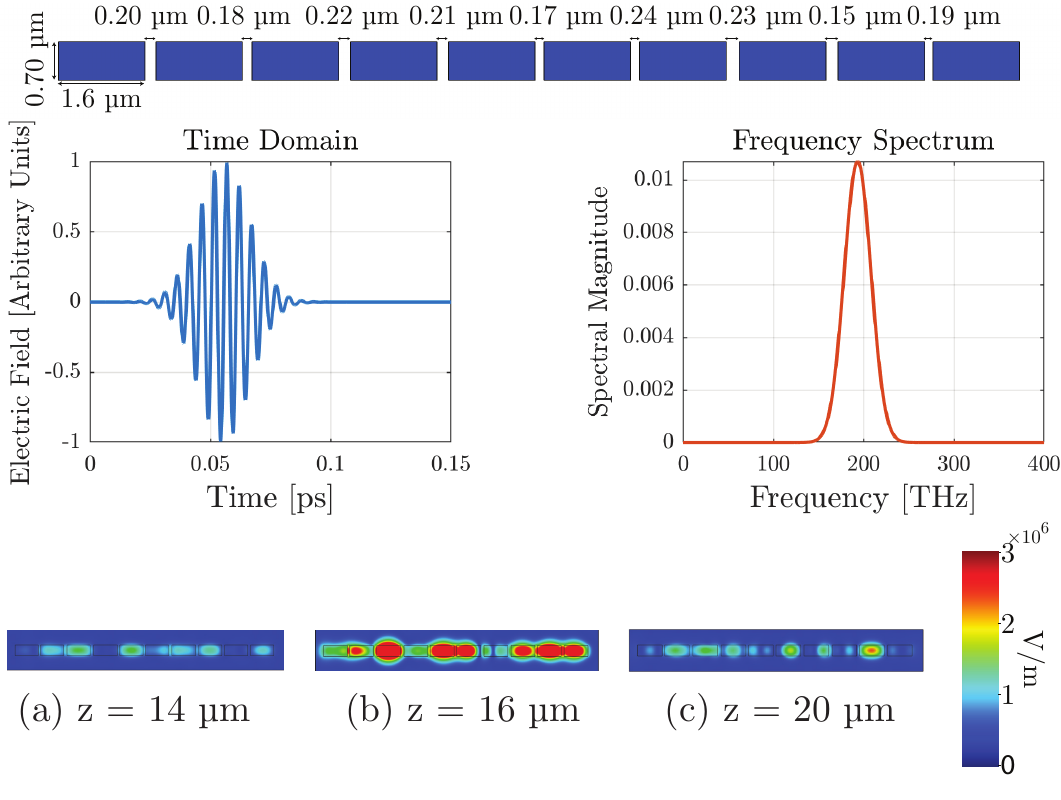}
\caption{First row:  schematic showing the cross section of a 3-D optical rectangular dielectric waveguide array. {Geometrical parameters and separation distance of ten identical waveguides are indicated. }The relative dielectric permittivities of the waveguides and the background medium are $\varepsilon_{1r}=3.9050$ and $\varepsilon_{sr}=2.0851$, respectively. Second row:  Gaussian pulse employed in the simulations with a wavelength 1.55 $\mu$m and its Fourier transform. { Broadband pulses are injected into the 2nd, 5th, 7th and 8th waveguides, which are arbitrarily chosen. 
Third row: cross-sectional electric field distributions in the entire system for the propagation distances: $z=14\mu$m, $z=16\mu$m and $z=20\mu$m at the fixed time $t=0.23$ps.}    } \label{Fields}
\end{figure} 

Here, a new scheme  for  space-time pulse retrieval is derived from the  differential equations governing the dynamics of  the complex amplitudes in coupled-mode waveguides which is applicable to  broadband pulses. A restoring mechanism is  identified  that  requires
an instantaneous (in time or space) relative phase shift  in  neighboring waveguides and a way to realize the scheme experimentally is suggested and verified  computationally.   
This is accomplished by performing full-wave numerical electromagnetic simulations for a coupled system of three-dimensional (3-D) dielectric rectangular waveguides.  Initial  broadband Gaussian pulses are injected in particular  waveguides and after passing some distance, the light distributes randomly in the coupled waveguides. This coupling distance and the amount of power distributed between the waveguides are controllable by engineering the effective coupling between the waveguides (cf. Fig. \ref{Pic}(b)). To decode  the signal a similar  structure (decoder) is used, but at the signal entry region a scatterer in every second waveguide causes  a phase shift. The numerical simulations confirm the analytical results (presented in the following subsection) that the output signal of the decoder faithfully images the input signal in the coder.

{\it Retrieval mechanism.} 
%The light intensity and the waveguide materials are assumed to be such that the wave propagation is in the linear regime.
For an injected pulse with the electric field vector $\omega$ frequency component ${\bf E}_\omega(x,y,z)$ the evolution  is governed by the  wave equation with spatial distribution of the relative dielectric permittivity of the composite  structure $\epsilon(x,y)+\delta\epsilon(x,y,z) $, where
$\epsilon(x,y)$  is the part which is  translationally invariant along the $z$-axis (propagation direction), whereas  $\delta\epsilon(x,y,z)$ can be an arbitrary   function of $x,y$ and $z$.
 As we are working in the linear regime, it is instructive to analyze the modes from which we construct the short pulses. An electric field with an angular frequency $\omega$ can be expanded in  a linear combination of normalized eigenmodes $\boldsymbol{\Phi}_\ell(x,y)$ of the isolated  waveguides ($\ell$ represents the number of waveguides): 
\begin{equation}
{\bf E}=\sum_\ell C_l(z){\boldsymbol \Phi}_\ell(x,y)e^{i(\omega t-k_\omega z)}+c.c.\ .
\label{phi}
\end{equation}
Here $k_\omega$ is a propagation constant, $C_\ell$ is a slowly varying amplitude of the mode in the $\ell$-th waveguide and it is a function of the propagation distance $z$. Following the procedure of coupled waveguide theory \cite{lederer}  one readily gets the discrete Schr\"odinger-type equation describing a stationary light distribution \cite{trompeter}:
\begin{equation}
2ik_\omega\frac{d C_\ell}{d
z}=J_{\ell}C_{\ell+1}+J_{\ell-1}C_{\ell-1}+V_\ell C_\ell.
\label{eq00}
\end{equation}
Here $J_\ell$ is   the coupling strength between the neighboring  $\ell$ and $\ell+1$ waveguides and the propagation constant  derives from 
$k_\omega^2=\frac{\omega^2}{c^2}\int \epsilon(x,y)|{\boldsymbol \Phi}_{\ell}|^2dxdy$, while the "potential" function is given by
$V_\ell(z)\equiv \frac{\omega^2}{c^2}\int \delta\epsilon(x,y,z)|{\boldsymbol \Phi}_{\ell}|^2dxdy$.

{ In \cite{szameit1,szameit2} the potential term has a staggered form produced by fragmented waveguide patterns which could be chosen as a delta function to demonstrate the effect: $V_\ell(z)=K(-1)^\ell\delta(z)$. For  the particular case:}
\begin{equation}
K=\pi k_\omega,
\label{K}
\end{equation}
and the unitarily transform via  $U$   
\begin{equation}
C_\ell=C^R_\ell U^\dagger =C^R_\ell  \exp\left[{-i\frac{\pi}{2}
(-1)^\ell\int\limits_{-\infty}^z\delta(s)ds}\right] .\label{eq01}
\end{equation}
gives \eqref{eq00} the following form:
\begin{equation}
2ik_\omega\frac{dC^R_\ell}{d z}=
\begin{cases}
J_{\ell}C^R_{\ell+1}+J_{\ell-1}C^R_{\ell-1}, \quad {\rm if\; }  z<0,\\
-\left(J_{\ell}C^R_{\ell+1}+J_{\ell-1}C^R_{\ell-1}\right), \quad {\rm if\; }  z>0.
\end{cases}
\label{eq03}
\end{equation}  
\begin{figure*}[t]
\centering
\includegraphics[width=0.84\linewidth]{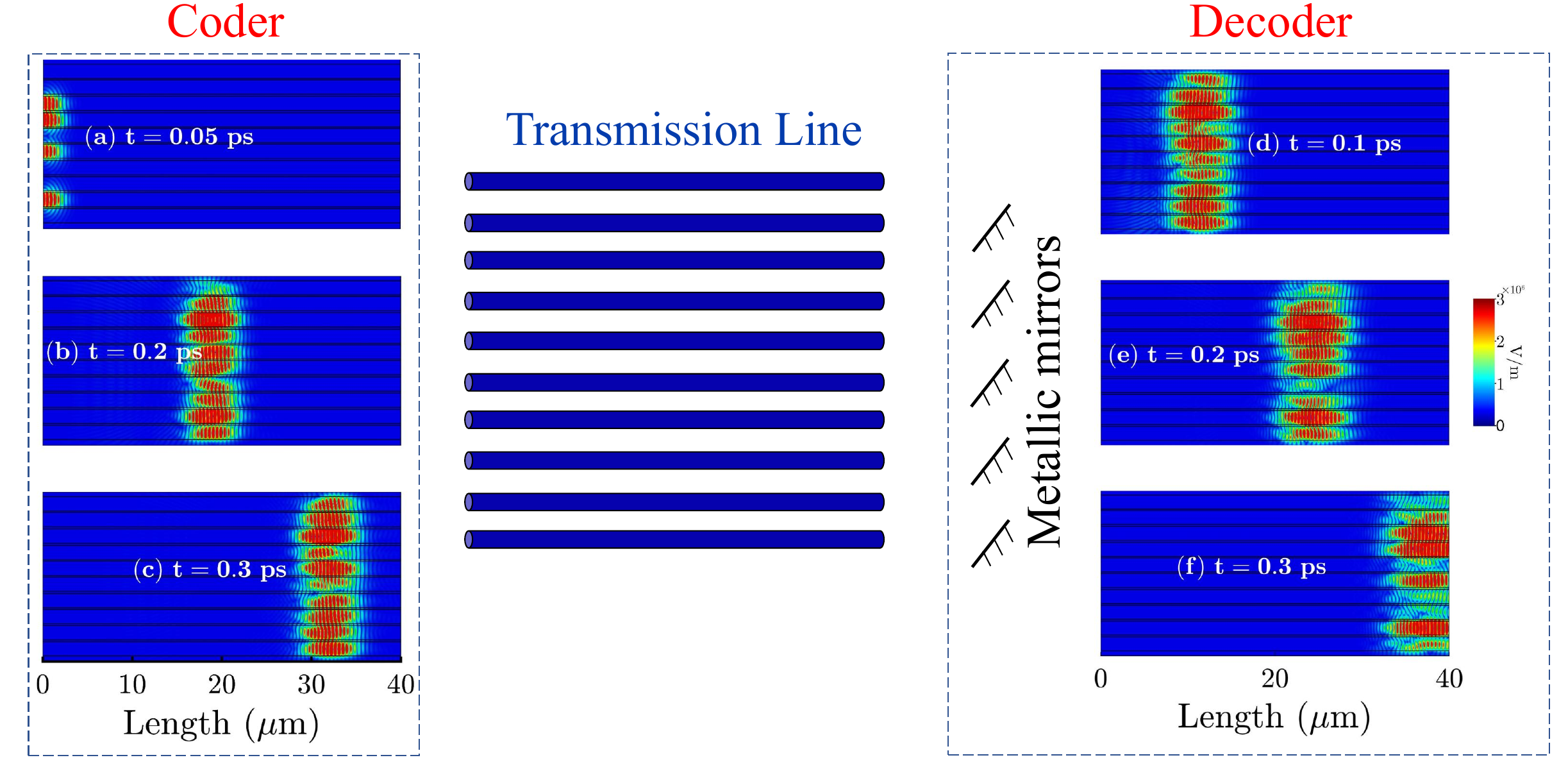}
\caption{Electric field  magnitude distribution of the signal propagation  in the waveguide array for the coder and decoder. { At $t=0$ initial Gaussian pulses are injected into the 2nd, 5th, 7th and 8th ports (c.f. Fig. 1).} The wavelength of the Gaussian pulse is $\lambda=1.55 \mu$m and the pulse duration is 18 fs. { The graphs (a)-(c) show the evolution of the signal for different times from $t=0.05$ps  to $t=0.3$ps in the coder.} In  practice, the encrypted  signal is decoupled and transmitted to the decoder via independent transmission lines. Phase-modifying metamaterials or perfectly reflecting mirrors, such as protected metallic (silver) mirrors, may be used to achieve the required phase shifts. { (d-f) show the signal decoding dynamics.}  }     \label{Time2}
\end{figure*}
\begin{figure}[h]
\includegraphics[width=0.9\linewidth]{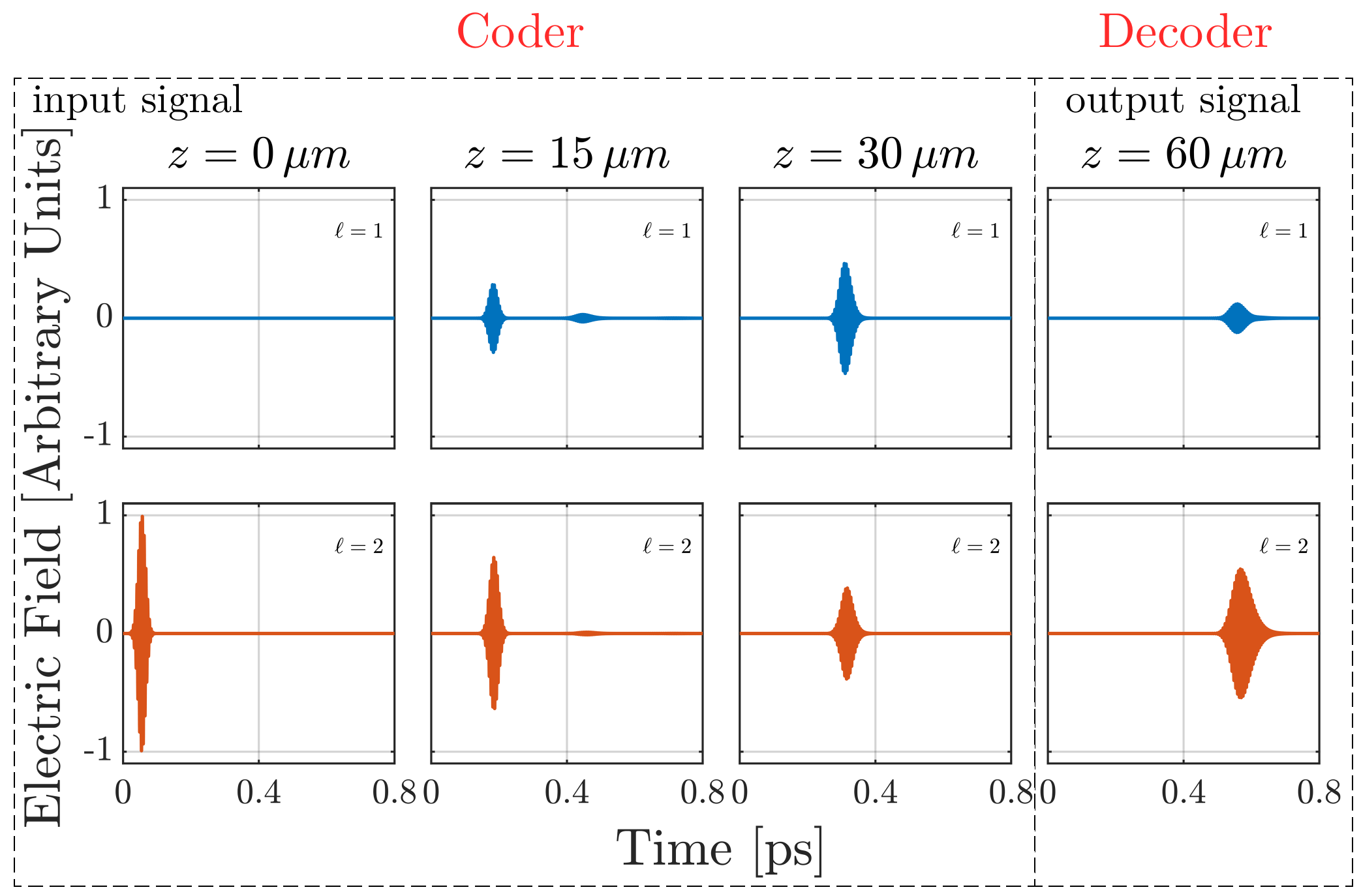} 
\caption{Time evolution of the pulsed input  signal at different cross sections in the individual waveguides labelled by  $\ell$.
Far right column shows the output signal from the decoder.}  \label{Plots}
\end{figure}
This relation implies that the evolution of the wave amplitude $C^R_\ell(z)$ for $z>0$ mirrors that for $z<0$ while the sign of propagation in the $z$-direction is flipped.
Thus, upon scattering from $V_\ell$, the propagation is coherently inverted such that  $C^R_\ell(z_0)=C^R_\ell(-z_0)$ allowing for a signal retrieval. 
%{This inversion mechanism is demonstrated in \cite{szameit1,szameit2} and holds for stationary beams.} 
For a broadband signal with a wide spectrum of propagation constants $k_\omega$, the condition \eqref{K} is however strongly violated and hence, a retrieval of the broadband signal is not possible. To remedy this situation, instead of considering a "potential" $V_\ell$, we apply at $z=0$ a phase shift of $\pi$  in odd numbered waveguides and for all harmonics of the pulses. Indeed, inspecting \eqref{eq00}, apart from $z=0$,
we can  write for different harmonics of a broadband signal the following expression:
\begin{equation}
2ik_\omega\frac{d C_\ell^\omega}{d
z}=J_{\ell}C_{\ell+1}^\omega+J_{\ell-1}C_{\ell-1}^\omega.
\label{eq08}
\end{equation}
Considering all harmonics, if  at $z=0$ the sign  of $C_\ell^\omega$  is inverted when  $\ell$ is  odd,  then the evolution is formally  reversed. 
An experimental realization of such a scenario could be, for example, phase-modifying  metamaterials attached/engineered appropriately at the input side of the decoder or perfectly reflecting  mirrors positioned at $z=0$.
Thus, one may envision an application as shown in Fig. \ref{Pic}:  a predefined broadband input signal injected in isolated waveguides
enters the coder (the section from $z = z_0$ to $z = l_c$), which is composed of an array of waveguides with possibly space-varying coupling strength.  Leaving the   coder at $z = l_c$,  the encrypted signal is transmitted via conventional isolated waveguides (transmission lines).  To decode the optical information,  at $z = l_d$, the signal passes the same distance in the decoder as in the coder (i.e., $l_d-l_t = l_c-z_0$) but after being mirror reflected at the entrance of odd numbered waveguides at $z = l_t$. Leaving the decoder, the cleared output signal propagates in  isolated fibers to the receiver.
The reliability of the scheme will be demonstrated in the following section for ultrashort pulses in coupled 3-D dielectric rectangular waveguides.

{\it Numerical simulations with pulsed signals.}
We demonstrate numerically a scheme for ten identical closely spaced 3-D rectangular dielectric waveguides (weak-coupling regime). The geometrical and material parameters of the waveguides, as well as the separation distances are summarized  in Fig. \ref{Fields}. We focus here only on the working principle of the coder and decoder. The pulse propagation 
in isolated waveguides is well documented  and  issues of dispersion suppression  and integration of optical fibers in photonic circuits  are established \cite{Weninger}.
The separation distances between the waveguides are different (resulting in a site dependent hopping constant $J_l$ in \eqref{eq00})    to accomplish  a random  signal at relatively short distances. The relative dielectric permittivities of the waveguides and the background medium are $\varepsilon_{1r}=3.9050$ and $\varepsilon_{sr}=2.0851$, respectively. The wavelength of the Gaussian pulse is  $\lambda=1.55\mu$m and a full width at half maximum (FWHM) of the pulse is about 0.015 ps. These short pulses are injected into the 2nd, 5th, 7th and 8th waveguides, which are arbitrarily chosen. The light tunnels into the neighboring waveguides due to coupling and at $z=40 \mu$m and one observes a completely mixed picture of the output signal, as evidenced by the left panel in Fig. \ref{Time2}.  Then, upon transferring the signal via independent optical fibers (simulations not shown), the signal is shifted in phase by $\pi$ only in the odd numbered waveguides keeping the phases for the even numbered waveguides unchanged.  As a result, after passing the decoder   (here, 40$\mu$m long) the initial broadband signal  is  retrieved (c.f.\, right panel in Fig. \ref{Time2}).   
\begin{figure}[t]
\includegraphics[width=1.0\linewidth]{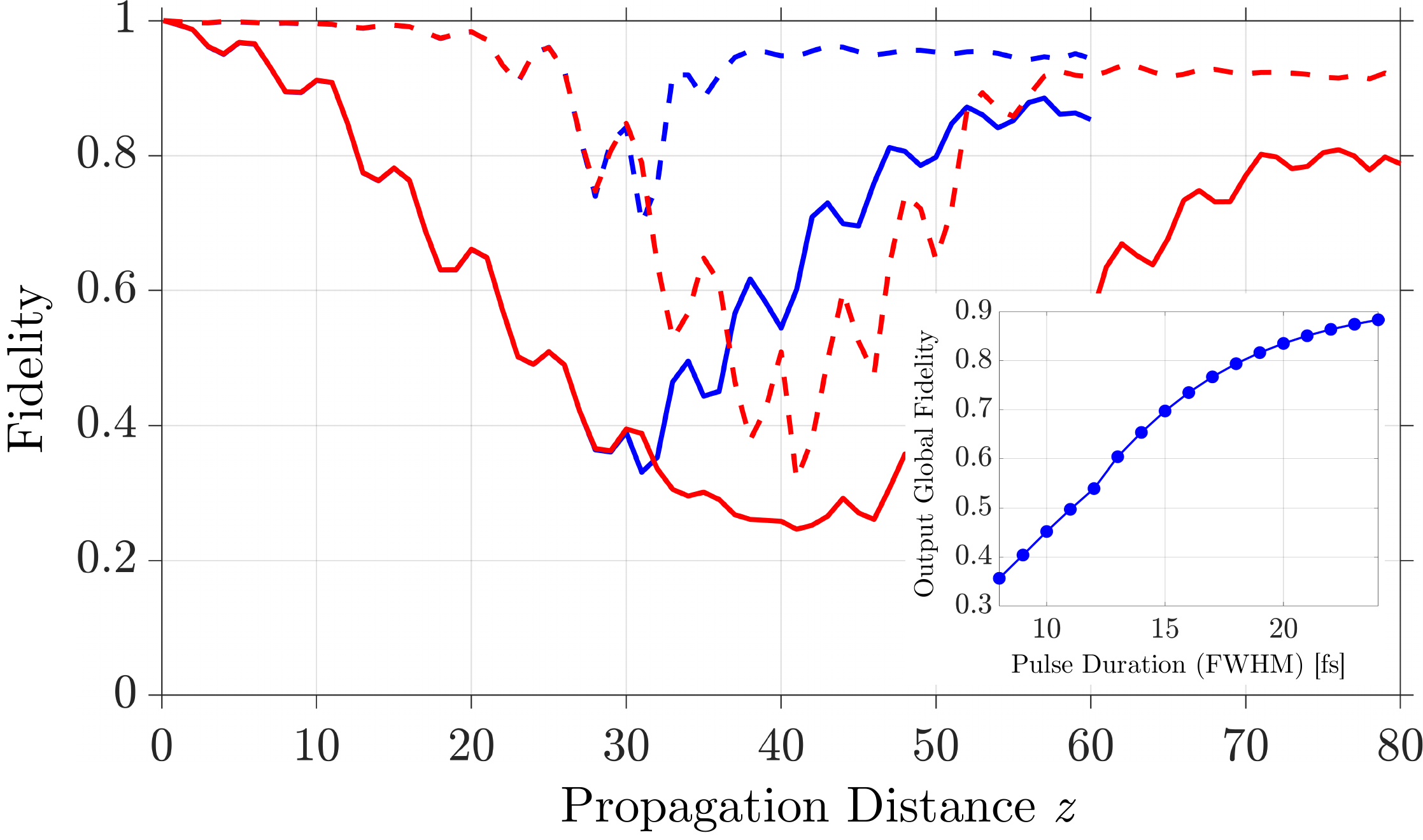} 
\caption{{ Global fidelity (solid lines) and phase fidelity (dashed lines) as a function of $z$ for two different lengths of the coder/decoder: $L = 30\mu$m (blue) and $L = 40\mu$m (red). Pulse duration is 18 fs. An inset shows a dependence of the output global fidelity on the pulse duration at $L = 40\mu$m.}}     \label{Power}
\end{figure}

For deeper insight,  the time evolution of the electric field is shown 
at different cross-sections of the waveguide arrays in the coder and decoder. Figure \ref{Plots} evidences that the signals are  retrieved at the end  of the decoder. The amplitudes are smaller than those of the injected pulse at $z = 0 \mu$m since the short pulses experience broadening due to dispersion inside the dielectric rectangular waveguide arrays. The dielectric rectangular waveguide generally causes strong pulse broadening. Nonetheless, the scheme is still useful
 even for this challenging case. { The pulse retrieval can be quantitatively characterized by the global fidelity ${\cal F}_{gl}$ and the phase fidelity ${\cal F}_{ph}$, respectively:
 \begin{eqnarray} \label{fidelity}
     {\cal F}_{gl}=\underset{\tau}{\text {max}}\frac{\bigl|\int {\bf E}_\omega(0)\cdot{\bf E}_\omega(z)e^{i\omega \tau}d\omega dxdy\bigr|^2}{\int  |{\bf E}_\omega(0)|^2d\omega dxdy\cdot\int |{\bf E}_\omega(z)|^2d\omega dxdy}; \nonumber \\ 
     {\cal F}_{ph}=\underset{\tau}{\text {max}}\frac{\bigl|\int |{\bf E}_\omega(0)|^2e^{i(\Phi_\omega(0)-\Phi_\omega(z)+\omega \tau)}d\omega dxdy\bigr|}{\int  |{\bf E}_\omega(0)|^2d\omega dxdy}
 \end{eqnarray}
 as displayed in Fig. \ref{Power}. 
 %Global fidelity incorporates both amplitude and phase information, thereby providing a complete characterization of the overlap between two states or wavefunctions. In contrast, phase fidelity isolates the phase component disregarding amplitude variations. 
 From the figure it follows that when the length of the coder/decoder is $40\mu m$ micrometers, the global fidelity at the end of the decoder is about 0.78, whereas it was 0.25 at the end of the coder. Additionally, we have studied a dependence of the output global fidelity on the pulse duration shown as an inset in Fig. 5. For the wider pulses, the global fidelity is approaching one. When the length of the coder/decoder is shorter, $L = 30\mu$m micrometers, the global fidelity at the end of the decoder is about 0.85 versus 0.35 at the end of the coder. The differences at the end of the decoder can be explained by the strong dispersion, which heavily broadens and deforms pulses.  An influence of pulse broadening may be reduced by employing structures with engineered dispersion. Particularly, photonic-crystal waveguides may be designed to provide a low group velocity dispersion (GVD).
 
To assess   the sensitivity of  retrieval  to  various imperfections in the coder/decoder  we assume  the lengths of the waveguides of decoding section are not equal and differ by  some value $\Delta L_{max}$, which varies from $0\mu m$ to $0.5\mu m$. The  signal retrieval coefficient, which is a ratio of a sum of the input and output powers, is displayed in Table I. When $\Delta L_{max}>0.2\mu m$ the signal retrieval is strongly compromised.
\begin{table}[h]
\centering
\caption{Signal retrieval (in percentage) as a function of the waveguide length differences of the decoding section.}
\begin{tabular}{|c|c|c|c|c|c|c|}
\hline
$\Delta L_{max}~(\mu m)$ & 0 & 0.1 & 0.2 & 0.3 & 0.4 & 0.5 \\ \hline
Rel. Recovery (\%) & 81.8 & 76.9 & 70.2 & 46.8 & 33.7 & 23.8 \\ \hline
\end{tabular} 
\end{table}
We  also investigated the influence of the mirror imperfections on the signal retrieval, keeping  fixed  the lengths of all waveguides and varying the reflection characteristics of the mirrors.  As realistic candidates for mirrors, we may consider ultrafast mirrors with low group delay dispersion (GDD) less than $30\text{ fs}^2$ operating in the wavelength region 1400 nm – 1700 nm and also protected metallic (silver) mirrors. The reflectance is about 99\% and the phase variation from $\pi$ across the bandwidth is about $5^{\circ}$. In order to assess the sensitivity of the proposed retrieval procedure to realistic mirror imperfections, we performed numerical simulations within $180^0 \pm 20^0$ range showing that these phase deviations do not lead to any appreciable degradation of the retrieval accuracy. }    

{\it Conclusions and future directions.} {For a coupled waveguide array, we demonstrated analytically a methodology for space-time retrieval of optical short pulses by introducing an instantaneous relative waveguide specific phase shift. Generalization of the problem to photonic devices  {\cite{jand1,jand2}} is  straightforward. { We are planning to study numerically planar waveguide arrays comprising aluminium gallium arsenide (AlGaAs) or silica glass; InP/InGaAsP quaternary substrate technology. The coupling length for photonic devices is much smaller, which will allow us to achieve the signal retrieval within a much smaller footprint.}

{\it Acknowledgment.} 
%D.E. and V.J. acknowledge the partial support by the Deutsche Forschungsgemeinschaft (DFG) in the Framework of the CRC/TRR 196 MARIE (Project‐ID 287022738) within project M03. 
R. Kh. acknowledges the STSM grant from Horizon Europe COST action CA23134. J.B. acknowledges DFG support through project Nr. 429194455. D.E. and V.J. also
acknowledge the partial support by the Deutsche Forschungsgemeinschaft (DFG) in the Framework of the CRC/TRR 196 MARIE (Project-ID
287022738) within project M03.  

{\it Disclosures.}
The authors declare no conflicts of interest.

\bibliography{Reviv}

\end{document}